# Detection of the Chiral Spin Structure in Ferromagnetic SrRuO$_3$ Thin Film

**Authors:**
H. Huang[1], S.-J. Lee[1], B. Kim[2,3,4,*], B. Sohn[2,3], C. Kim[2,3], C.-C. Kao[5], and J.-S. Lee[1,*]

**Affiliations:**
[1]*Stanford Synchrotron Radiation Lightsource, SLAC National Accelerator Laboratory, Menlo Park, California 94025, USA*

[2]*Department of Physics and Astronomy, Seoul National University, Seoul 08826, S. Korea*

[3]*Center for Correlated Electron Systems, Institute for Basic Science, Seoul 08826, S. Korea*

[4]*Department of Energy Science, Sungkyunkwan University, Suwon, 16419, S. Korea*

[5]*SLAC National Accelerator Laboratory, Menlo Park, California 94025, USA*

*Correspondence to: bongju@skku.edu; jslee@slac.stanford.edu

**A SrRuO$_3$ (SRO) thin film and its heterostructure have brought much attention because of the recently demonstrated fascinating properties, such as topological Hall effect and skyrmions. Critical to the understanding of those SRO properties is the study of the spin configuration. Here, we conduct resonant soft x-ray scattering (RSXS) at oxygen *K*-edge to investigate the spin configuration of a 4 unit-cell SRO film that was grown epitaxially on a single crystal SrTiO$_3$. The RSXS signal under a magnetic field (~0.4 Tesla) clearly shows a magnetic dichroism pattern around the specular reflection. Model calculations on the RSXS signal demonstrate that the magnetic dichroism pattern originates from a Néel-type chiral spin structure in this SRO thin film. We believe that the observed spin structure of the SRO system is a critical piece of information for understanding its intriguing magnetic and transport properties.**

A ferromagnetic oxide SrRuO$_3$ (SRO) thin film has attracted wide research interest because of their remarkable physical properties and potential spintronics applications.[1-7] The controllable properties of SRO thin films through interface engineering or electric field gating have especially fascinated researchers.[7-9] For example, the SRO based heterostructures with the heavy metal layer (e.g. SrRuO$_3$/SrIrO$_3$) have been additionally highlighted due to the demonstrations of both magnetic quasiparticles, i.e. skyrmion, within its structure and the topological Hall effect.[8-11] This is because topologically non-trivial magnetic skyrmions are important ingredients to make spintronics devices with high stability.[12-16] In the SrRuO$_3$/SrIrO$_3$ case, the broken inversion symmetry at the interface together with the large spin-orbit coupling from the heavy metal layer lead to interfacial Dzyaloshinskii-Moriya interaction (DMI) in SRO.[12,13,17-19] The DMI effect induces the chiral spin structure and forms into skyrmion quasiparticles.

Meanwhile, it was recently suggested that the single-layered SRO film (i.e., without the heavy metal capping layer) is possible to show DMI due to the oxygen octahedron distortion at the interface between SRO and a substrate.[20-22] In this sense, a chiral spin structure is expected even in this single-layered SRO thin film as well, which makes it a powerful candidate material for spintronics devices. As a consequence, many transport measurements have been performed on the single-layered SRO thin films in search of a chiral spin structure signature. In the transverse Hall resistance on the SRO film, a hump-like feature has been observed in a certain temperature and field range. In terms of the interpretation of the feature, however, there is a controversy. Some studies have regarded the hump-like feature as the topological Hall effect originating from the chiral spin phase,[20, 21, 23] as it has been observed in magnetic skyrmion systems.[24-26] Other studies attribute it to the inhomogeneous property of ultrathin SRO films.[27,28] Thus, these contrary interpretations suggest that the Hall effect measurement is insufficient to determine the spin configuration in SRO thin film.[23]

In this study, we employed a resonant soft x-ray scattering (RSXS) approach to study the spin structure in a 4 unit-cell (u.c.) SRO film grown on a SrTiO$_3$ (STO) substrate. This is because the RSXS experiment was recently established to be a powerful non-destructive technique for determining the chiral spin structure in both magnetic multilayer and bulk skyrmion systems.[29-34] We observed a magnetic circular dichroism pattern around the specular peak at the oxygen *K*-edge. Through model calculations on the dichroic RSXS pattern, we revealed an existence of the Néel-type chiral spin structure in the SRO thin film. We believe that this finding is a crucial achievement in understanding its magnetic properties, paving the way for future spintronics applications even on the single-layered SRO system.

High-quality SRO thin (4 u.c.) film was epitaxially grown on STO substrate using the pulsed laser deposition method.[23] Figure 1(a) schematically shows the atomic structure of the SRO/STO. According to the previous studies,[4,35,36] this 4 u.c. SRO exhibits the ferromagnetic behavior at low temperature. Using the resistivity measurement of our thin film as a function of temperature as well as its partial derivative (Figure 1(b)), we checked that the ferromagnetic transition occurs at ~76.3 K. And besides, we checked the Hall resistance measurements. As shown in Figure 1(c), we could see the clear a bump feature at 20 K with varying external magnetic field (more results at different temperature are shown in Figure S2), which is beyond the ordinary Hall resistance and anomalous Hall resistance. As aforementioned, such the bump feature in Hall resistivity is possibly attributed to the topological chiral spin structure,[20-23] but still elusive.[27,28]

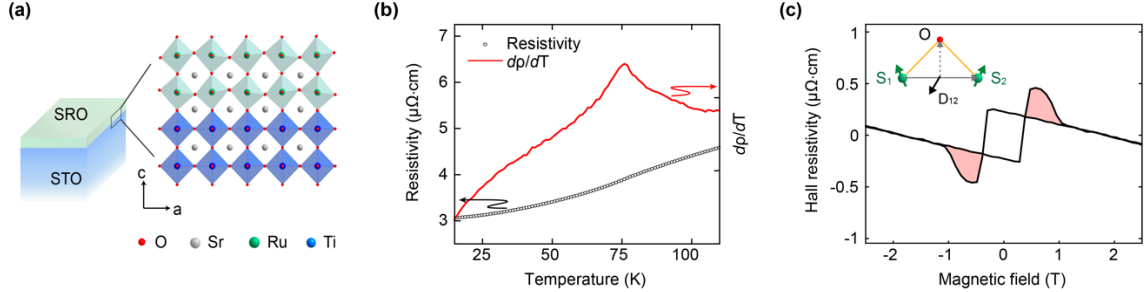

**Figure 1.** Thin $SrRuO_3$ (4 u.c.) film on the single crystal $SrTiO_3$. (a) Schematic cartoon of the film structure. The distorted arrangement of SRO's octahedral around the interface region is highlighted. (b) The temperature dependence of the resistivity (black colored) and its partial derivative (red). It shows a transition around $T = 76.3$ K. (c) The transverse Hall resistivity of the SRO film measured at $T = 20$ K. The colored shades denote the additional term of Hall resistivity. The inset shows the effect of DMI between two neighboring Ru spins ($S_1$ and $S_2$) and the central oxygen.

Since lattice mismatch exists between the epitaxial SRO film (3.93 Å) and STO substrate (3.905 Å), the oxygen octahedral distortion is accompanied around the interfacial layer of SRO (Figure 1(a)). This octahedral distortion breaks the inversion symmetry at the interface, contributing to the DMI.[20-22] Specifically, the oxygen (O) atom between two neighboring ferromagnetic Ru spins ($S_1$ and $S_2$) departures from the line connecting two Ru atoms due to the octahedral distortion (the inset of the Figure 1(c)). The displacement of the central oxygen determines the strength of the Dzyaloshinskii vector ($D_{12}$), and the interfacial DMI equals to $D_{12} \cdot (S_1 \times S_2)$ in this SRO film.[38] In other words, the participation of O atoms is essential for the DMI, and also the magnetic structure in SRO thin film.

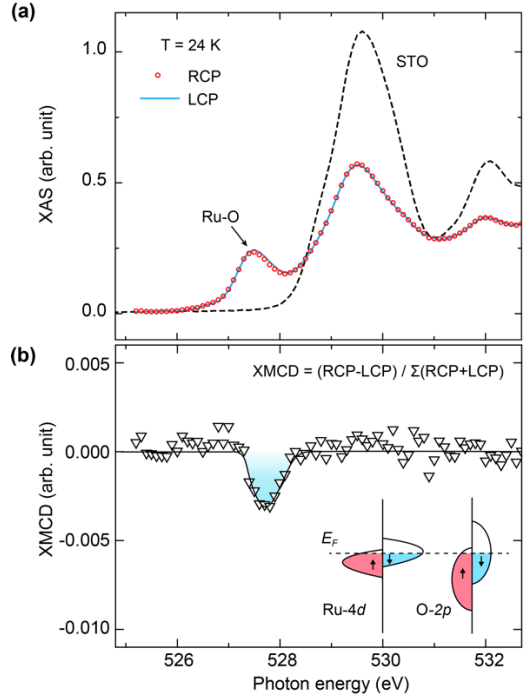

**Figure 2.** X-ray spectroscopy measurements on the SRO (4 u.c.)/STO. (a) X-ray spectroscopy measurements at O $K$-edge XAS spectra with RCP (red-circles) and LCP (solid-line). The peak at $E_{ph}$ = 527.5 eV corresponds to the Ru 4$d$-O 2$p$ bonding state. As a reference, a pure STO spectrum from the substrate material only is superposed (dashed line). (b) The XMCD at the O $K$-edge of the SRO thin film. The blue colored shade in the XMCD denotes the magnetic dichroism signal of the Ru-O bonding state. The inset depicts the ferromagnetically polarized band of oxygen band via the hybridization with that of Ru one.

To explore a role of the oxygen in this single-layered SRO, we firstly performed x-ray magnetic circular dichroism (XMCD) measurement at O $K$-edge with a sample temperature of $T$ = 24 K which is well below the measured ferromagnetic transition temperature, $T_c$ ~ 76.3 K. A magnetic field (~0.4 Tesla) is applied along the surface-normal direction. This is because the thin SRO film undergoes the perpendicular magnetic anisotropy[38] accompanied by the weakened dipole interaction.[39] Figure 2(a) shows the x-ray absorption spectroscopy (XAS) spectra of SRO thin film measured by the incident right/left circular polarized (RCP/LCP) x-rays. We observed a small, but clear difference of the XAS spectra between RCP and LCP around 527.5 eV, where is the Ru-O bond assigned via the comparison with the reference STO XAS spectrum. The difference (i.e., XMCD) between RCP and LCP spectra is presented in Figure 2(b). We found the XMCD signal around the incident x-ray photon energy, $E_{ph}$ = 527.8 eV. This result indicates that the oxygen 2$p$ band is magnetically polarized through the strong hybridization between Ru-4$d$ band and O-2$p$ bond (the inset in Figure 2(b)).

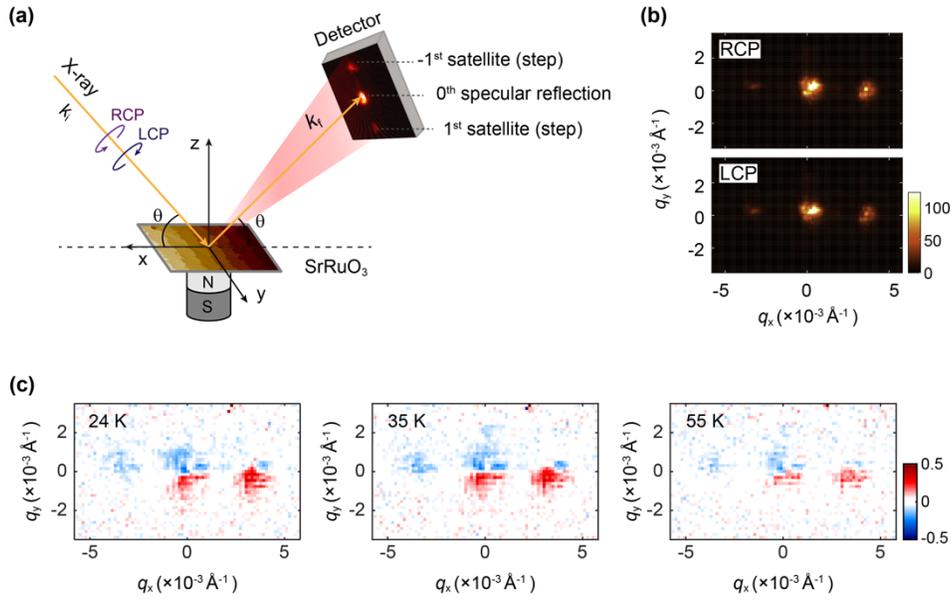

**Figure 3.** Resonant soft x-ray scattering on the SRO (4 u.c.) film. (a) Schematic picture of the RSXS experimental geometry and detected pattern. $k_i$, $k_f$ denote the incident (75°) and outgoing (150°) x-ray, respectively. Along the sample normal direction, the magnetic field (~ 0.4 Tesla) is applied. The scattered x-rays are detected by an area detector (CCD). The sample surface is represented by atomic force microscopy (AFM) image. It shows the step-terrace pattern (see also Fig. S1). (b) The scattering patterns with the RCP and LCP condition (measured at $T$ = 35 K). Reflections around $q_x$ = 0 and ~ ±0.004 Å$^{-1}$ are the specular (0$^{th}$) and satellite (±1$^{st}$) peaks, respectively. (c) The magnetic asymmetry pattern, $A = (I_{RCP} - I_{LCP})/(I_{RCP} + I_{LCP})$, at three different temperatures after subtracting the $A$-pattern at 79 K as a nonmagnetic background.

Next, we performed the RSXS experiment at 527.8 eV where the oxygen magnetic dichroism maximized, aiming to explore the spin structure of the SRO/STO. Figure 3(a) shows a schematic experimental configuration. In a similar logic to the XMCD measurement, a magnetic field was applied along the normal direction and the incident scattering angle ($\theta$) was maximally set at 75º to ensure a larger magnetic scattering signal under the magnetic field. A large area detector ($2\theta$) was placed at 150º to collect the spatial distribution of the scattering signals. We observed a bright specular peak (i.e., $2\theta/2 = \theta$) on the center of the detector as well as a series of satellite peaks along the $q_x$ direction, which is originated from the interference around the vicinal step terrace of the STO substrate (Figure 3(a) and Figure S1).

Considering the recent studies which demonstrated the magnetic dichroic pattern around the specular,[31,32] the polarization dependent RSXS is a powerful method to explore the spiral spin structure. We measured the specular peak with different circular polarized x-ray, i.e., RCP and LCP. As shown in Figure 3(b), there is a subtle but noticeable difference around the specular reflection between two polarizations. To extract such the difference, we plotted the magnetic asymmetry ratio, $A = (I_{RCP} - I_{LCP})/(I_{RCP} + I_{LCP})$, after removal of high temperature (above $T_c$) background (see details in Figure S3), where $I_{RCP}$ and $I_{LCP}$ are the RSXS signals of with the RCP and LCP, respectively. Figure 3(c) shows an evident magnetic dichroic signal around the specular reflection. The magnetic asymmetry pattern ($A$-pattern) is separated into two parts along the $q_y$ direction, with different sign of contrast. Note that although the dichroic signal is not conclusive at the satellite peak, the overall shape resembles that of the specular reflection. In particular, the shape of SRO's $A$-pattern is very similar to that of the recently reported $A$-pattern in an Ir/Co/Pt multilayer, which is originated from the Néel-type domain walls.[31,32] These findings imply the existence of chiral spin structure in the single-layered SRO system.

Furthermore, this implication is supported by the temperature dependence of which this *A*-pattern of the SRO is more evident at the lower temperature and getting dilute when the sample temperature is approaching to its $T_c$. Although, this chiral spin structure is detected via the O's magnetic resonance, it can be used to represent the spin configuration in $SrRuO_3$ system because of the strong hybridization of the O and Ru bands.

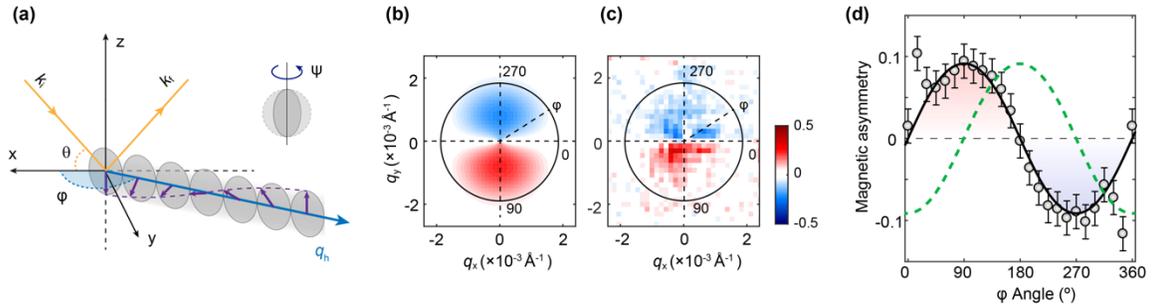

**Figure 4.** Theoretical analysis of the dichroic RSXS pattern. (a) A modeled spin spiral structure, $q_h$ is the spin propagation wavevector, φ represents the azimuth angle of $q_h$ rotated from the x-axis. $\Psi$ represents the rotation of the common rotation plane. (b) The simulated *A*-pattern from Néel-type spin structure. The in-plane pattern is coordinated with an azimuthal angle φ. (c) The experimental *A*-pattern (measured at 35 K) around the specular peak region is highlighted. (d) Azimuthal dependence of the integrated intensity along the radial direction in *A*-pattern. The experimental data (circles) is compared with the calculated Néel- (black solid-line) and Bloch-type (green dashed-line) spin structures. The error bar is estimated from the zero-field magnetic asymmetry pattern.

In order to scrutinize the details of the observed chiral spin structure in the single-layered SRO, we executed model calculations which simulate RSXS signals based on the magnetic scattering theory.[40] We started the scattering simulation from a Néel-type spin configuration which is typically induced by the interfacial DMI. Figure 4(a) schematically shows the modeled Néel-type spin configuration, in which $k_i$ and $k_f$ represent the incident and outgoing x-ray, respectively. The spiral spin is propagating along the $q_h$ direction, where $q_h$ is the wavevector of the spin modulation. The angle $\varphi$ represents the azimuth angle of $q_h$ rotated from the *x*-axis in *x-y* plane. Along the $q_h$ direction, the individual spins rotate or precess in a common rotation plane, as indicated by the gray circles in Figure 4(a). By changing the common rotation plane angle $\psi$, the spin spiral structure with difference helicity (e.g., Néel-type, Bloch-type, and hybridized-type) can be considered in this simulation. Note another the representative structure of Bloch-type is discussed in the Supplementary information. In this circumstance, the magnetic moment of the spin spiral structure is described as follows: $\boldsymbol{M} = (m_1, m_2, m_3)$, where $m_1 = M_0 \sin\psi \sin(\boldsymbol{q}_h \cdot \boldsymbol{r})$; $m_2 = -M_0 \cos\psi \sin(\boldsymbol{q}_h \cdot \boldsymbol{r})$; $m_3 = M_0 \cos(\boldsymbol{q}_h \cdot \boldsymbol{r})$. With the change of circular polarization of incident x-ray, the magnetic asymmetry ratio at $q_h$ around the specular peak, $I^\varphi(\boldsymbol{q}_h)$, is calculated as follows (see more details in Supporting Information):

$$I^\varphi(\boldsymbol{q}_h) = \frac{2\cos\theta \sin^2\theta \cos(\varphi + \psi)}{\sin^2\theta + \cos^2\theta \sin^2(\varphi + \psi) + 2\sin^2\theta \cos^2\theta \cos^2(\varphi + \psi)}.$$

Figure 4(b) shows the *A*-pattern simulated from the Néel-type spin structure obtained from the equation above. The shape of the simulated pattern is in good agreement with the experimental results (Figure 4(c)). Such an agreement is further supported by a comparison of the integrated magnetic asymmetry as a function of $\varphi$, as shown in Figure 4(d). The experimental result is well matching with the simulation from the Néel-type structure. If the helicity of the spin structure changes, the magnetic asymmetry curves will have a different phase from the experimental results. As indicated in Figure 4(d), the simulated result from Bloch type shows a 90° phase difference with the experimental result (see Figure S4b). Therefore, these results unambiguously confirmed the existence of Néel-type spin structure in SRO thin film.

It is finally worthy to note that this kind of the observed Néel-type spin structure in the SRO thin film would be possibly originated from the chirality of magnetic domain walls, or small quasiparticle magnetic structures (e.g. magnetic skyrmion) in SRO thin film.[10,31] Regardless of its origin, our result bears great significance in that the single-layered SRO thin film has a remarkable chiral spin structure even without assistance of other heavy metal layers. This chiral spin configuration may also help to irradiate the underlying mechanism for the novel magnetic properties which has been widely observed in SRO system.[24-26]

In summary, we have comprehensively studied the single-layered SRO thin film (4 u.c.) grown on an STO substrate using resonant soft x-ray scattering at the oxygen *K*-edge under 0.4 Tesla magnetic field. The RSXS signal with the magnetic field clearly shows a magnetic dichroism pattern around the specular reflection. Corresponding the model calculations on the magnetic pattern reveal the Néel-type spin structure in the thin SRO film. We believe that this result provides a critical clue to settle the controversial interpretations on the transport result on SRO and eventually an important fundamental ingredient to understand the magnetic properties in SRO thin film. Ultimately, it sheds new light on the potential of making spintronics devices even in such a simple-structured SRO film.

**Methods**

***Sample growth and characterization.*** The SRO thin films were grown by the pulsed laser deposition (PLD) method on the $TiO_2$- terminated STO substrate along [001] direction. The substrate was pre-annealed at 1300 K for 90 minutes to achieve a sharp terrace surface. A 248 nm KrF excimer laser with a power density of 1-2 $J/cm^2$ was used to deposit the SRO film at 1000 K and in 100 mTorr $O_2$. The film growth is monitored by an in-situ RHEED system. The flat surface quality was confirmed by the atomic force microscopy (AFM) measurement (Figure S1). The transport property was measured by the Physical Property Measurement System (PPMS), Quantum Design Inc.

***Resonant Soft X-ray Scattering and X-ray Magnetic Circular Dichroism.*** The RSXS and XMCD measurements were carried out at the Beamline 13-3 in Stanford Synchrotron Radiation Lightsource (SSRL), SLAC National Accelerator Laboratory. In both measurements, a permanent magnet was used to apply magnetic field (~ 0.4 Tesla) along the film normal direction. This is because the easy axis of magnetization is out-of-plane direction. The incident beam direction is aligned to the maximized $\boldsymbol{k_i} \cdot \boldsymbol{M}$ (where $\boldsymbol{k_i}$ is the wavevector of incident x-ray and $\boldsymbol{M}$ is the magnetic moment), leading to the large $\theta$ angle (close to normal direction). For XMCD measurement, the circular polarization (RCP and LCP) is controlled with the fixed field direction. All spectra were acquired by the total electron yield (TEY) mode. The RSXS experiment is done at specular reflection geometry with an incident angle of $\theta = 75°$ (detector angle, $2\theta = 150°$). The diffracted x-rays were collected by an area detector (CCD) at the specular position. The sample is mounted on an in-vacuum four-circle diffractometer and cooled down through an open cycle liquid helium system (the lowest temperature can be reached in this set-up is ~24 K).

***Theoretical analysis of RSXS signals.*** Using the same geometry as the experiment, the scattering intensity is calculated by $I(\bm{q}) \propto |\sum_n f_n \exp(i\bm{q} \cdot \bm{r}_n)|^2$. The structure factor of the magnetic scattering term is determined by the polarization of the incident and scattered x-ray and the magnetic moment. The magnetic structure with different helicity is modeled as discussed in the main text. When $\psi = \pm 90°$, the normal direction of a common rotation plane is perpendicular to the $\bm{q}_h$ direction, which is the case of a pure Néel-type structure, as shown in Figure 4(a). On the other hand, when $\psi = 0$ or $180°$, the normal direction of a common rotation plane is parallel to the $\bm{q}_h$ direction, representing the pure Bloch-type structure (see Figure S4). When $\psi$ is in-between, the model represents a hybrid chiral spin structure of the Néel- and Bloch- type. The magnetic asymmetry ratio $I^\varphi(\bm{q}_h)$ is obtained by changing the helicity of the incident beam from RCP to LCP. It represents the magnetic asymmetry ratio around the specular peak with a wavevector of $\bm{q}_h$. The simulated *A*-pattern is obtained by giving the Gaussian broadening of the magnetic asymmetry ratio along the radial direction, in which the Gaussian peak center is estimated from the experimental scattering peak center $\bm{q}_h$. The more details are described in the Supporting Information.

## Supporting Information

It is available from the corresponding author upon a request.

**Acknowledgments.** All soft x-ray experiments (scattering and spectroscopy) were carried out at the SSRL (beamline 13-3), SLAC National Accelerator Laboratory, supported by the U.S. Department of Energy, Office of Science, Office of Basic Energy

Sciences under Contract No. DE-AC02-76SF00515. B. Kim, B. Sohn, and C. Kim acknowledge the support from the research program of Institute for Basic Science (Grant No. IBS-R009-G2).**References**

(1) G. Koster, L. Klein, W. Siemons, G. Rijnders, J. S. Dodge, C.-B. Eom, D. H. A. Blank, M. R. Beasley, Structure, physical properties, and applications of SrRuO$_3$ thin films. *Rev. Mod. Phys.* **2012**, *84*, 253

(2) H. Boschker, T. Harada, T. Asaba, R. Ashoori, A. V. Boris, H. Hilgenkamp, C. R. Hughes, M. E. Holtz, L. Li, D. A. Muller, H. Nair, P. Reith, X. Renshaw Wang, D. G. Schlom, A. Soukiassian, J. Mannhart, Ferromagnetism and Conductivity in Atomically Thin SrRuO$_3$. *Phys. Rev. X* **2019**, *9*, 011027.

(3) S. Kang, Y. Tseng, B. H. Kim, S. Yun, B. Sohn, B. Kim, D. McNally, E. Paris, C. H. Kim, C. Kim, T. W. Noh, S. Ishihara, T. Schmitt, J.-G. Park, Orbital-selective confinement effect of Ru 4d orbitals in SrRuO$_3$ ultrathin film. *Phys. Rev. B* **2019**, *99*, 045113.

(4) J. M. Rondinelli, N. M. Caffrey, S. Sanvito, N. A. Spaldin, Electronic properties of bulk and thin film SrRuO$_3$: Search for the metal-insulator transition. *Phys. Rev. B* **2008**, *78*, 155107.

(5) Y. Ou, Z. Wang, C. S. Chang, H. P. Nair, H. Paik, N. Reynolds, D. C. Ralph, D. A. Muller, D. G. Schlom, R. A. Buhrman, Exceptionally High, Strongly Temperature Dependent, Spin Hall Conductivity of SrRuO$_3$. *Nano Lett.* **2019**, *19*, 3663.

(6) D. Kan, R. Aso, R. Sato, M. Haruta, H. Kurata, Y. Shimakawa, Tuning magnetic anisotropy by interfacially engineering the oxygen coordination environment in a transition metal oxide. *Nat. Mater.* **2016**, *15*, 432.

(7) B. Pang, L. Zhang, Y. B. Chen, J. Zhou, S. Yao, S. Zhang, Y. Chen, Spin-Glass-Like Behavior and Topological Hall Effect in SrRuO$_3$/SrIrO$_3$ Superlattices for Oxide Spintronics Applications. *ACS Appl. Mater. Inter.* **2017**, *9*, 3201.

(8) Y. Ohuchi, J. Matsuno, N. Ogawa, Y. Kozuka, M. Uchida, Y. Tokura, M. Kawasaki, Electric-field control of anomalous and topological Hall effects in oxide bilayer thin films. *Nat. Commun.* **2018**, *9*, 213.
13